\documentclass[twocolumn,showpacs,preprintnumbers,amsmath,amssymb,showkeys]{revtex4}

\usepackage{epsfig}
\usepackage[utf8]{inputenc}
\usepackage[T1]{fontenc}

\usepackage{graphicx}
\usepackage{dcolumn}
\usepackage{bm}
\usepackage{color}
\usepackage{slashed}

\newcommand{\cb}{\overline c}

\begin{document}

\title{Two-point correlation functions of QCD in the Landau gauge}

\author{Marcela Pel\'aez$^{a,b}$}

\author{Matthieu Tissier$^a$}

\author{Nicol\'as Wschebor$^b$}

\affiliation{\vspace{.2cm}
$^a$LPTMC, Laboratoire de Physique Th\'eorique de la Mati\`ere 
Condens\'ee, CNRS UMR 7600, Universit\'e Pierre et Marie Curie, \\
bo\^\i te 121, 4 
place Jussieu, 75252 Paris Cedex 05, France.\\
$^b$Instituto de F\'{\i}sica, Facultad de Ingenier\'{\i}a, Universidad de la Rep\'ublica,\\
J.~H. y Reissig 565, 11000 Montevideo, Uruguay.}

\date{\today}

\begin{abstract}
  We investigate the gluon, ghost and quark propagators in the Landau
  gauge with dynamic quarks. We perform a one-loop calculation in a
  model where the standard Faddeev-Popov Lagrangian is complemented by
  a mass term for the gluons which is seen as a minimal way of taking
  into account the effect of the Gribov copies.  The analytic results
  are compared with lattice data obtained in four dimension and for
  two, three and four quark flavors. The gluon and ghost propagators
  are reproduced with a few percent accuracy in the whole range of
  accessible momenta. The scalar part of the quark propagator is found
  to be in good agreement with the lattice data. However, the quark
  renormalization is poorly described. We attribute this discrepancy
  to the fact that the one-loop corrections to this quantity are
  unusually small so that the two loop contribution can not be
  discarded. The results are expressed in terms of the coupling, the
  gluon mass and the light quark mass at 1 GeV.

\end{abstract}

\pacs{12.38.-t, 12.38.Aw, 12.38.Bx,11.10.Kk}

\maketitle

\section{Introduction}

\label{sec_intro}

It is well-known that the Faddeev-Popov construction, which is the
standard analytic method for fixing the gauge, is not sufficient in order
to treat the infrared regime of QCD. The problem originates in the existence
of Gribov copies~\cite{Gribov77} which are ignored in
the Faddeev-Popov construction. Gribov copies do not play any role in
the ultraviolet regime. For that reason, the Yang-Mills action
complemented by the Faddeev-Popov action is a good starting point to
analyze both gauge-invariant and non-invariant quantities at momentum
scales much bigger than $1$ GeV.  However, at low momenta the question
of how to include the effect of Gribov copies in a Lagrangian still
remains unsolved even though some ideas have been already developed.
Among those, the most popular is probably the Gribov–Zwanziger
proposal \cite{Zwanziger89,Zwanziger92,Zwanziger01,Dudal08}. It
consists in restricting the functional integral to the first Gribov
region by adding several new fields.  Unfortunately it was shown in
\cite{vanBaal:1991zw} that the first Gribov region also includes many
Gribov copies so that the Gribov ambiguity is not completely
removed. Moreover, the Gribov-Zwanzinger procedure relies on some
formal manipulations that are not fully justified from first
principles.

In the past decades, lattice simulations improved considerably and
became the most reliable technique to describe the infrared region of
QCD \cite{montvay94}. In particular, quenched lattice simulations in
Landau gauge
\cite{Sternbeck:2007ug,Cucchieri_08b,Cucchieri_08c,Sternbeck:2008mv,Cucchieri09,Bogolubsky09,Dudal10} and unquenched
simulations \cite{Bowman:2004jm,Parappilly:2006si}
have demonstrated that the gluon propagator acquires a finite value at
low momentum, contrary to the original belief. Numerical simulations
succeeded in convincing the community that, in this gauge, the gluon
propagator behaves in the IR as if it was massive.

Even before that, the infrared regime was studied through several
semi-analytical methods, among which the more popular were based on
Dyson-Schwinger (DS) equations. Depending on the precise implementation,
two solutions were observed, called scaling and decoupling
(or massive) solutions. The scaling solution is characterized by a
gluon propagator that vanishes as a power law at low momentum
\cite{vonSmekal97,Alkofer00,Zwanziger01,Fischer03,Bloch03,Fischer08}.
On the other hand, the decoupling solution corresponds to a finite
gluon propagator
\cite{Aguilar04,Boucaud06,Aguilar07,Aguilar08,Boucaud08,Fischer08,RodriguezQuintero10,Huber:2012kd},
in agreement with lattice results.

Acknowledging the fact that no analytic method can deal with the
Gribov ambiguity in a fully consistent way, two of the authors have
proposed an alternative strategy to study analytically the IR regime
of the theory. The idea relies on working with the simplest Lagrangian
which reproduces lattice results.  To account for the decoupling
solution observed on the lattice in a minimal way, the Faddeev-Popov
Lagrangian is complemented by a mass term for the gluons. In euclidean
space, the Lagrangian density reads:
\begin{equation}
\label{eq_cf}
  \begin{split}
\mathcal{L}= &\frac{1}{4}F_{\mu\nu}^aF_{\mu\nu}^a +\partial_{\mu}\bar
c^a(D_{\mu}c)^a+ih^a\partial_{\mu}A_{\mu}^a \\
   &+\frac{m^2}{2}A_{\mu}^aA_{\mu}^a+\sum_{i=1}^{N_f} \bar\psi_i(\gamma_\mu D_\mu + M_i)\psi_i,    
  \end{split}
  \end{equation}
  where $g$ is the coupling constant, $\gamma_\mu$ are euclidean Dirac
  matrices satisfying $\{\gamma_\mu,\gamma_\nu\}=2\delta_{\mu,\nu}$,
  the flavor index $i$ runs over the $N_f$ quark flavors and
\begin{align*}
F_{\mu\nu}^a&=\partial_{\mu}A_{\nu}^a-\partial_{\nu}A_{\mu}^a+gf^{abc}A_{\mu}^bA_{\nu}^c,\\
(D_{\mu}c)^a&=\partial_{\mu}c^a+gf^{abc}A_{\mu}^bc^c,\\
D_{\mu}\psi&=\partial_{\mu}\psi-igA_{\mu}^a t^a \psi.
\end{align*}
The latin indices correspond to the
$SU(N)$ gauge group, $t^a$ are the generators of the algebra in the
fundamental representation and $f^{abc}$ are the structure constants.

The model described by Eq.~(\ref{eq_cf}) is a particular case of the
Curci-Ferrari model \cite{Curci76}. It is well known that the mass
term violates the BRST symmetry of the Faddeev-Popov
Lagrangian. However, it is still symmetric under a modified BRST
symmetry responsible for its renormalizability.  This Lagrangian can
also be motivated from first principles by taking into account Gribov
copies \cite{Serreau:2012cg}.

A great advantage of the phenomenological model described by
Eq.~(\ref{eq_cf}) is that it is very simple and allows to perform
perturbative calculations very easily. Moreover, we found
\cite{Tissier:2010ts,Tissier:2011ey} that there exist renormalization
schemes in which the coupling constant remains finite at all
momentum scales. The absence of Landau pole allows us to implement perturbation
theory even in the infrared regime. The quenched one-loop calculations
for the two-point \cite{Tissier:2010ts,Tissier:2011ey} and three-point
\cite{Pelaez:2013cpa} correlation functions compare very well with
lattice simulations. The previous model was also used for studying the
two-point correlation functions at finite temperature
\cite{reinosa14} where it reproduces at a qualitative level the
properties of the gluon and ghost propagators.

In this article, we pursue our systematic comparison of the
correlation functions obtained within the model (\ref{eq_cf}) with
those extracted from lattice simulations. We include here dynamical
quarks and compute, at one loop, the two-point correlation function
for the gluon, ghost and quark, for arbitrary dimension, number of
colors ($N$) and flavors ($N_f$). The rest of the article is organized
as follows. In Sect. \ref{glgh}, we discuss the 1-loop calculation for
the gluon and ghost propagators which are expressed in terms of
Passarino-Veltman integrals \cite{Passarino78}. We present our results
in arbitrary dimension and give explicit expressions in the physically
relevant case $d=4$.  Sect. \ref{quark} is devoted to the quark
propagator. In Sect. \ref{sec_renorm} we introduce the renormalization
schemes and discuss the implementation of the renormalization
group. In Sect. \ref{sec_res} we perform a comparison of lattice
correlation functions with our unquenched perturbative results in the
gluon and ghost sector. We focus on $N=3$ with two light degenerates
quarks ($N_f=2$) and with two light quarks and two heavier quarks
($N_f=2+1+1$).  In Sect. \ref{sectionNf21} we work with $N_f=2+1$. The
quark propagator was extracted from lattice simulations in this case,
which enables us to make a comparison with our analytical results.
Finally, in Sect. \ref{sec_error} we estimate the two-loop
contributions which gives an indication of the error bars on our
results.

\section{Unquenched Gluon and Ghost propagator}
\label{glgh}
The gluon and ghost propagators have been extensively studied in
lattice simulations, both in the quenched and unquenched case. It is
found that the addition of the sea quarks does not change qualitatively
the behavior of both propagators. It however tends to lower the plateau
observed at small momenta for the gluon propagator.

We parametrize the gluon and ghost 2-point vertex functions in the
following way
\begin{align}
&  
\Gamma^{(2)}_{A_\mu^aA_\nu^b}(p)=\delta^{ab}\big[\Gamma^\perp(p)P^\perp_{\mu\nu}(p)
+\Gamma^\parallel(p)P^\parallel_{\mu\nu}(p)\big],\\
& \Gamma^{(2)}_{c^a \cb^b}(p)=\delta^{ab} \frac{p^2}{J(p)}, 
\end{align}
where $P^\perp_{\mu\nu}$ and $P^\parallel_{\mu\nu}$ are the transverse
and parallel projector respectively, defined as:
\[P^\parallel_{\mu\nu}(p)=\frac{p_\mu
  p_\nu}{p^2}\hspace{5mm}\text{and}\hspace{5mm}P^\perp_{\mu\nu}(p)=\delta_{\mu\nu}-P^\parallel_{\mu\nu}(p).\]

In linear covariant gauges and in particular in the linear version of the Landau gauge without the inclusion
of the mass term, the gluon self-energy is transverse. Here, the longitudinal part is non-zero but it is
controlled by a non-renormalization theorem (see  \cite{Tissier:2011ey}). It is important to stress, however, that this longitudinal part has
no impact on the gluon propagator, which is transverse in Landau gauge even in presence of the mass term. Similar remarks apply to higher vertex 
functions, that contains longitudinal parts, but which do not contribute to the corresponding correlation functions.
The function $J(p)$ (usually called the dressing function of the
ghost) and the $\Gamma^\perp(p)$ (the transverse part of the two-point
gluon vertex, related to the gluon dressing function $\frac{p^2}{\Gamma^\perp(p)}$) explicitly appear in the ghost and gluon propagators
and are therefore of special interest in order to compare our
calculations with lattice results. As the longitudinal part of the
two-point gluon vertex is not directly accessible in lattice
simulations, we do not describe it here.
 
\begin{figure}[htbp]
\includegraphics[width=2cm]{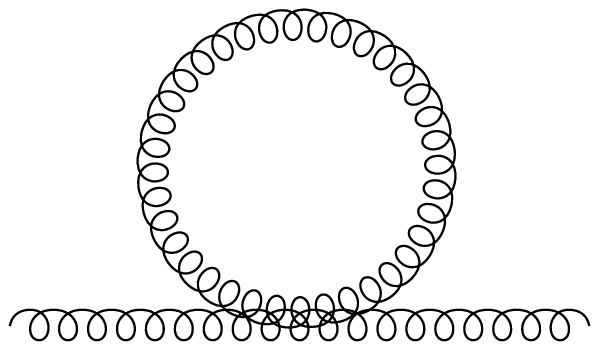}
\includegraphics[width=2cm]{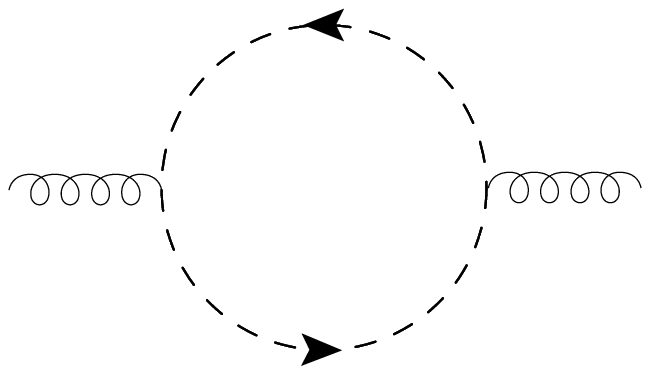}
\includegraphics[width=2cm]{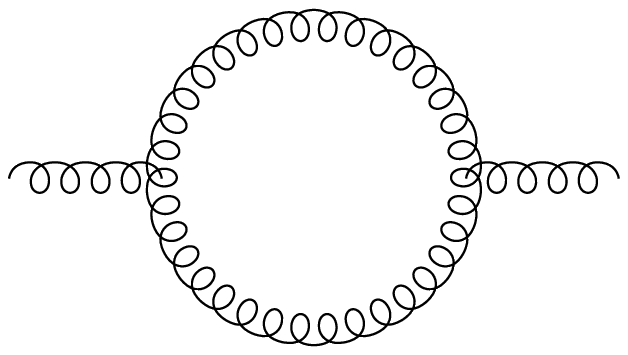}
\includegraphics[width=2.1cm]{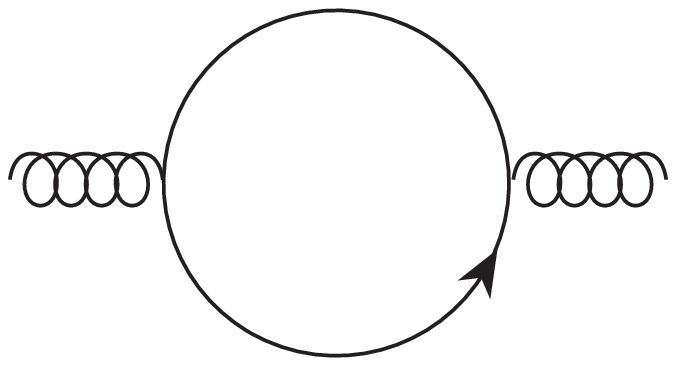}

\vspace{.8cm}
\includegraphics[width=2.5cm,angle=0]{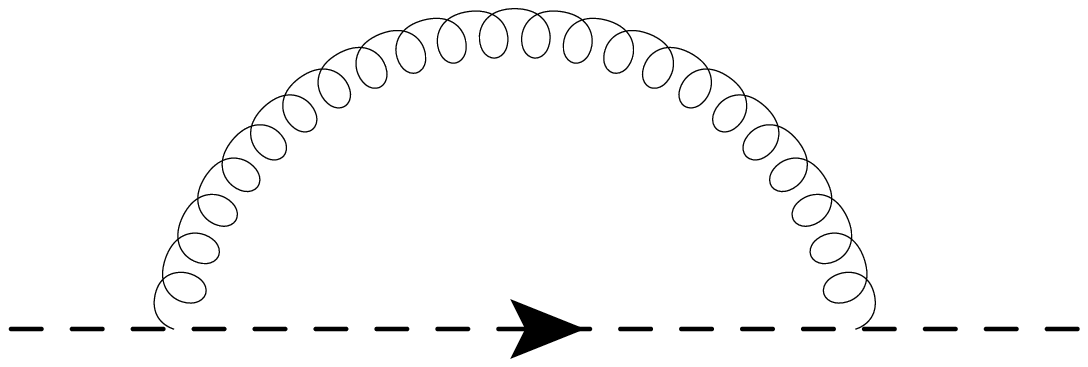}
\caption{\label{diagrams}First line: four diagrams contributing to the gluon self-energy. Second line: diagram contributing
to the ghost self-energy}
\end{figure}

At one-loop, the gluon and ghost two-point vertex functions are given
by the diagrams shown in Fig.~\ref{diagrams}.
Most of these diagrams were already computed in
\cite{Tissier:2010ts}. In what concerns the gluon
propagator, we just need here to compute the diagram with a quark loop
(fourth diagram of Fig.~\ref{diagrams}), which can be expressed in
terms of Passarino-Veltman integrals as
\begin{equation}
\label{eq_diag_quark_loop}
\begin{split}
\Gamma_{\text{1loop},4}^\perp(p) = & \frac{2 g^2 T_f}{(d-1)} \left\{[4 M^2-(d-2) p^2]
   B_0(p,M,M)\right. \\
& \left. +2 (d-2) A(M)\right\},
\end{split}
\end{equation}
where the index $4$ represents the fourth diagram in
Fig.~\ref{diagrams} and $T_f$ is defined by ${\rm Tr}(t^at^b)=T_f
\delta^{ab}$ (in the fundamental representation, $T_f=1/2$). The $A$
and $B_0$ functions are the analogue of Passarino-Veltman
integrals~\cite{Passarino78} in euclidean space:
\[A(m_1)=\int \frac{d^dq}{(2\pi)^d} \frac{1}{q^2+m_1^2} \]
\[B_0(p,m_1,m_2)=\int \frac{d^dq}{(2\pi)^d}
\frac{1}{q^2+m_1^2}\frac{1}{(q+p)^2+m_2^2} \] Each quark flavor
contributes to the gluon vertex and it is therefore necessary to sum
this diagram over the flavors, with $M$ replaced by the corresponding
quark mass. We have checked that (\ref{eq_diag_quark_loop}) coincides
with the expression of~\cite{davydychev96} when the quark mass is set
to zero.

In $d=4-\epsilon$, the diagram can be expressed in a completely
analytical form:
\begin{equation*}
\begin{split}
\Gamma_{\text{1loop},4}^\perp&(p) = \frac{
  g^2 T_f p^2}{6 \pi ^2}\Big\lbrace-\frac{1}{\epsilon}+\log \left(\frac{M e^{\gamma/2}}{\sqrt{4 \pi}}\right)- \frac{5}{6} +2 t\\
& +\left(1-2 t\right) \sqrt{4 t+1} \coth^{-1}\left(\sqrt{4
    t+1}\right)\Big\rbrace+\mathcal O(\epsilon)\\
\end{split}
\end{equation*}
where $t=\frac{M^2}{p^2}$.

At one loop, the ghost propagator is the same as in the quenched
situation due to the non-existence of a ghost-quark vertex. However,
the ghost dressing function will be indirectly influenced by the quarks
through the renormalization-group flow of the coupling constant and
gluon mass.

\section{Quark propagator}
\label{quark}
\begin{figure}[htbp]
\includegraphics[width=4cm,angle=0]{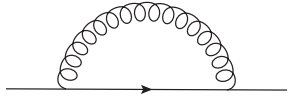}
\caption{\label{quarkdiagram} diagram contributing to the quark two point vertex}
\end{figure}

The one-loop contribution to the quark two-point vertex involves only
one diagram which is shown in Fig.~\ref{quarkdiagram}. It has two
independent structures in Dirac indices and is therefore parametrized
by two scalar functions:
\[\Gamma^{(2)}_{\psi\bar\psi}(p^2)= Z^{-1}(p)\left(i\slashed p + M(p)\right)\]
Both $M(p)$ and $Z(p)$ have been determined in lattice simulations.

This diagram is expressed in terms of Passarino-Veltman integrals as follows:
\begin{equation*}
\begin{split}
\Gamma^{(2)}_{\psi\bar\psi,\text{1loop}}&(p^2)=  g^2 (d-1) M
B_0(p,m,M)\\
&-i\slashed{p} \frac{g^2 C_f}{2 m^2 p^2}  \Big\lbrace\big[(2-d)
m^4+(d-3) m^2 (M^2-p^2) \\
&+(M^2+p^2)^2\big]B_0(p,m,M)\\&+(M^2+p^2)^2 B_0(p,0,M)
+A(m)[(2-d)m^2\\
&-M^2-p^2]+(d-2) m^2 A(M)\Big\rbrace,
\end{split}
\end{equation*}
where $C_f$ is defined by $t^at^a=C_f \openone$. (In the fundamental
representation, $C_f=\frac{N^2-1}{2N}$.)  The previous expression
coincides with that of \cite{davydychev00} when the gluon mass $m$ is
set to zero.

In $d=4-\epsilon$, the diagram takes the analytical form:
\begin{align*}
\Gamma^{(2)}_{\psi\bar\psi,\text{1loop}}&(p^2) = \frac{-i\slashed{p} g^2 C_f}{64 \pi ^2 m^2 p^4}\Big\lbrace k^2 \left[2 m^4+m^2(p^2-M^2)\right.\\
& \left.-(M^2+p^2)^2\right] Q -2 m^2 p^2(-2 m^2+M^2 +p^2) \\
&-2[2 m^6+3 m^4(p^2-M^2)+(M^2+p^2)^3] \log \left(\frac{M}{m}\right)\\
&  -2 (M^2+p^2)^3 \log \left(\frac{M^2+p^2}{M^2}\right)\Big\rbrace\\
& +\frac{3 g^2 C_f M}{8 \pi ^2}\Big\lbrace -\frac{1}{
   \epsilon}+\log\left(\frac{m e^{\gamma/2}}{\sqrt{4 \pi} }\right)-\frac{2}{3}\\
& -\frac{k^2}{4 p^2} Q +\frac{1}{2 p^2} (m^2-M^2+p^2) \log
   \left(\frac{M}{m}\right)\Big\rbrace\\
& \\
&+ \mathcal O(\epsilon)
\end{align*}
with:
\begin{equation}
  \label{eq_kq}
  \begin{split}
    k^2&=\sqrt{m^4+2 m^2 (p^2-M^2)+(M^2+p^2)^2},\\
    Q&=\log
    \left[\frac{(k^2-p^2)^2-(M^2-m^2)^2}{(k^2+p^2)^2-(M^2-m^2)^2}\right].
  \end{split}
\end{equation}

It is worth to be mentioned that, at one loop, the field
renormalization (the part proportional to $\slashed p$ in the previous
expression) is finite. In fact, it even vanishes in the limit of
vanishing gluon mass. 

\section{Renormalization and Renormalization Group}
\label{sec_renorm}
\subsection{Renormalization scheme}
In four dimensions most of the expressions presented above are
divergent. In order to absorb these divergences, we redefine the
coupling constant, masses and fields by introducing renormalization
factors:
\begin{align*}
 A_0^{a\,\mu}=& \sqrt{Z_A} A^{a\,\mu},
 c_0^{a}= \sqrt{Z_c} c^{a},
 \psi_0^{a}= \sqrt{Z_\psi} \psi^{a},
 \bar c_0^{a}= \sqrt{Z_c} \bar c^{a}, \nonumber\\
g_0&= Z_g g \hspace{.5cm} m_0^2= Z_{m^2} m^2 \hspace{.5cm} M_0= Z_M M
\end{align*}
The index $0$ represents the bare quantities and for now on, when
not specified, all quantities are the renormalized ones. 

The renormalization constants are redefined in the infrared-safe (IS) scheme: 
\begin{align}
\label{rencond}
&\Gamma^\perp(p=\mu)=m^2+\mu^2, \hspace{.4cm} J(p=\mu)=1,\nonumber\\
& Z(p=\mu)=1, \hspace{.4cm} M(p=\mu)=M,\nonumber\\
&Z_{m^2} Z_A Z_c=1, \hspace{.4cm} Z_g\sqrt{Z_A} Z_c=1.
\end{align}
The IS scheme is convenient because it does not present a Landau pole
\cite{Tissier:2011ey}. It combines the Taylor scheme~\cite{taylor71}
and a non-renormalization theorem for the gluon mass, conjectured in
\cite{Gracey:2002yt}, and proved in
\cite{Dudal02,Wschebor07,Tissier08}.

The explicit expressions for the renormalization factors are given in
\cite{Tissier:2011ey} for $N_f=0$ and for arbitrary $N_f$ their
divergent parts match in Landau gauge with the results presented in
\cite{Gracey:2002yt}. 

\subsection{Renormalization Group}

The renormalization procedure leads to finite expressions for the
vertex functions. However, as is well known, these expressions are
hampered by large logarithms and we have to implement the
renormalization-group procedure to control perturbation theory. The
$\beta$ functions and anomalous dimensions of the fields are:
\begin{align*}
\beta_g(g,m^2,\{M_i\})&=\mu\frac{dg}{d\mu}\Big|_{g_0, m^2_0,M_{i,0}},\\
\beta_{m^2}(g,m^2,\{M_i\})&=\mu\frac{dm^2}{d\mu}\Big|_{g_0, m^2_0,M_{i,0}},\\
\gamma_A(g,m^2,\{M_i\})&=\mu\frac{d\log Z_A}{d\mu}\Big|_{g_0, m^2_0,M_{i,0}},\\
\gamma_c(g,m^2,\{M_i\})&=\mu\frac{d\log Z_c}{d\mu}\Big|_{g_0, m^2_0,M_{i,0}},\\
\beta_{M_i}(g,m^2,\{M_i\})&=\mu\frac{dM_i}{d\mu}\Big|_{g_0, m^2_0,M_{i,0}},\\
\gamma_{\psi_i}(g,m^2,\{M_i\})&=\mu\frac{d\log Z_{\psi_i}}{d\mu}\Big|_{g_0, m^2_0,M_{i,0}}.
\end{align*}

For completeness, we give here the contribution of the quarks to the
various $\beta$ and $\gamma$ functions: 
\[\gamma_A^{\text{quarks}}=\sum_{i=1}^{N_f}\frac{g^2 T_f}{\pi^2} \Big\lbrace\frac{t-6}{6 t}-\frac{2 \log \left(\frac{\sqrt{t+4}-\sqrt{t}}{\sqrt{t}+\sqrt{t+4}}\right)}{\sqrt{t^3 (t+4)}}\Big\rbrace\]
\begin{align*}
\gamma_\psi &=
\frac{g^2 C_f}{16 \pi ^2 m^2 \mu^4} \Big\lbrace
m^2 \mu^2 \left(4m^2-2 M^2+ \mu^2\right) \\ 
& +\big[-2 \left(2 m^6-3 m^4 M^2+M^6\right)-3 \mu^2 \left(M^4+m^4\right)+\mu^6\big]\\
&\times \log \left(\frac{M}{m}\right)+\left(-2 M^6-3 M^4 \mu^2+\mu^6\right) \log
   \left(\frac{M^2+\mu^2}{M^2}\right)\\
&+\frac{1}{2 k^2}\left[2(m^2-M^2)^3 (2 m^2+M^2)\right.\\
&+\mu^2(m^2-M^2)(7m^4+6 m^2M^2+5 M^4)\\
& \left.+\mu^4 \left(3 m^4-2 m^2 M^2-3 M^4\right)+ \mu^6 \left(m^2+M^2\right)+\mu^8\right]Q\Big\rbrace
\end{align*}
\begin{align*}
\beta_M &=M\gamma_\psi + \frac{3 g^2 C_fM}{8 \pi^2} \left\lbrace -u^2 +(m^2-M^2) \log \left(\frac{M}{m}\right)\right.\\
&\left .-\frac{1 }{2 k^2}\left(m^4+m^2 \left(u^2-2 M^2\right)+M^2 \left(M^2+u^2\right)\right)Q\right\rbrace
\end{align*}
where $Q$ is the expression appearing in Eq.~(\ref{eq_kq}) with $p$
replaced by the renormalization-group scale $\mu$. As discussed above,
the ghost 2-point function is not affected by the quarks at one loop
so that $\gamma_c$ is not modified and we recall that, in the IS
scheme, $\beta_{m^2}=m^2(\gamma_A+\gamma_c)$ and
$\beta_g=g(\gamma_A/2+\gamma_c)$.

We can then use the RG equation for the vertex function with $n_A$
gluon legs, $n_c$ ghost legs and $n_\psi$ quark legs:
\begin{equation*}
\begin{split}
\Big( \mu \partial_\mu -&\frac 1 2 (n_A \gamma_A+n_c \gamma_c+n_\psi \gamma_\psi)\\&+\beta_g 
\partial_{g}+
\beta_{m^2}\partial_{m^2}+
\sum_i\beta_{M_i}\partial_{M_i}\Big)\Gamma^{(n_A,n_c,n_\psi)}=0,
\end{split}
\end{equation*}
to relate these functions at different scales, giving:
\begin{equation*}
\label{eq_int_RG}
\begin{split}
\Gamma&^{(n_A,n_c,n_\psi)}(\{p_i\},\mu,g(\mu),m^2(\mu),\{M_i(\mu)\})=\\
&z_A(\mu)^{n_A/2}z_c(\mu)^{n_c/2}z_\psi(\mu)^{n_\psi/2}\\
&\times\Gamma^{(n_A,n_c,n_\psi)}(\{p_i\},\mu_0,g(\mu_0),m^2(\mu_0),\{M_i(\mu_0)\}).
\end{split} 
\end{equation*} 
Here $g(\mu)$, $m^2(\mu)$ and $M_i(\mu)$ are obtained by integration of the
beta functions with initial conditions given at some scale $\mu_0$ and:  
\begin{equation*}
\label{eq_def_z_phi}
\begin{split}
\log z_A(\mu)&=\int_{\mu_0}^\mu\frac
     {d\mu'}{\mu'}\gamma_A\left(g(\mu'),m^2(\mu'),M(\mu')\right),\\ \log
     z_c(\mu)&=\int_{\mu_0}^\mu\frac
     {d\mu'}{\mu'}\gamma_c\left(g(\mu'),m^2(\mu'),M(\mu')\right),\\ \log
     z_\psi(\mu)&=\int_{\mu_0}^\mu\frac
     {d\mu'}{\mu'}\gamma_\psi\left(g(\mu'),m^2(\mu'),M(\mu')\right).
\end{split}
\end{equation*}
Note that each quark mass $M_i$ has its own $\beta_{M_i}$ function
that must be integrated. In our 1-loop calculation, the flow of $M_i$
does not depend on all the quark masses but only on $M_i$ itself. As the
infrared safe scheme does not present a Landau pole the RG scale $\mu$
will be chosen as $\mu\simeq p$ for a correlation function with
typical momentum $p$ in order to avoid large logarithms.

\section{Results for the gluon and ghost sectors}
\label{sec_res}

In this section we compare the gluon and ghost propagator obtained in
our one-loop analytical expressions and in lattice simulations,
for $SU(3)$ with $N_f=2$ and $N_f=2+1+1$.

When comparing our findings with the lattice data, we have to fix the
initial conditions of the renormalization-group flow. These quantities
were considered here as fitting parameters and were chosen to minimize
simultaneously the relative error for the gluon and ghost propagators,
respectively defined as:
\begin{align}
\label{eq_chi}
\chi^2_{AA}&=\frac{1}{N}\sum_i\Gamma_{\rm lt.}^\perp(p_i)^2\left(\frac{1}{\Gamma_{\rm 
lt.}^\perp(p_i)}-\frac 1{\Gamma_{\rm th.}^\perp(p_i)}\right)^2\nonumber\\
\chi^2_{c\cb}&=\frac{1}{N}\sum_i J^{-2}_{\rm lt.}(p_i)
\left(J_{\rm lt.}(p_i)-J_{\rm th.}(p_i)\right)^2\nonumber
\end{align}
There are only few parameters to choose in the fitting procedure. They
correspond to the initial conditions at some scale $\mu_0$ of the
coupling constant and masses in the renormalization-group flow. In the
$N_f=2+1+1$ case there are {\it a priori} three quark masses to
fit. However we fixed, in the initial condition, the middle and heavy
quarks to be twice and 20 times heavier than the lightest one
(to obtain the same mass ratios as obtained from the lattice at two GeV). Therefore we have
to fit only three parameters: the coupling constant, the gluon mass
and the light quark mass, all at some renormalization scale (we choose
$\mu_0=1$~GeV here).\footnote{The propagators are, generally speaking,
not normalized as in (\ref{rencond}). In consequence, by using multiplicative renormalizability, we
multiplied each lattice propagator by a constant in order to impose the renormalization condition (\ref{rencond}).
In the case of unrenormalized propagators, this also implements the required field renormalization.}

\begin{figure}[htbp]
 \includegraphics[width=\linewidth]{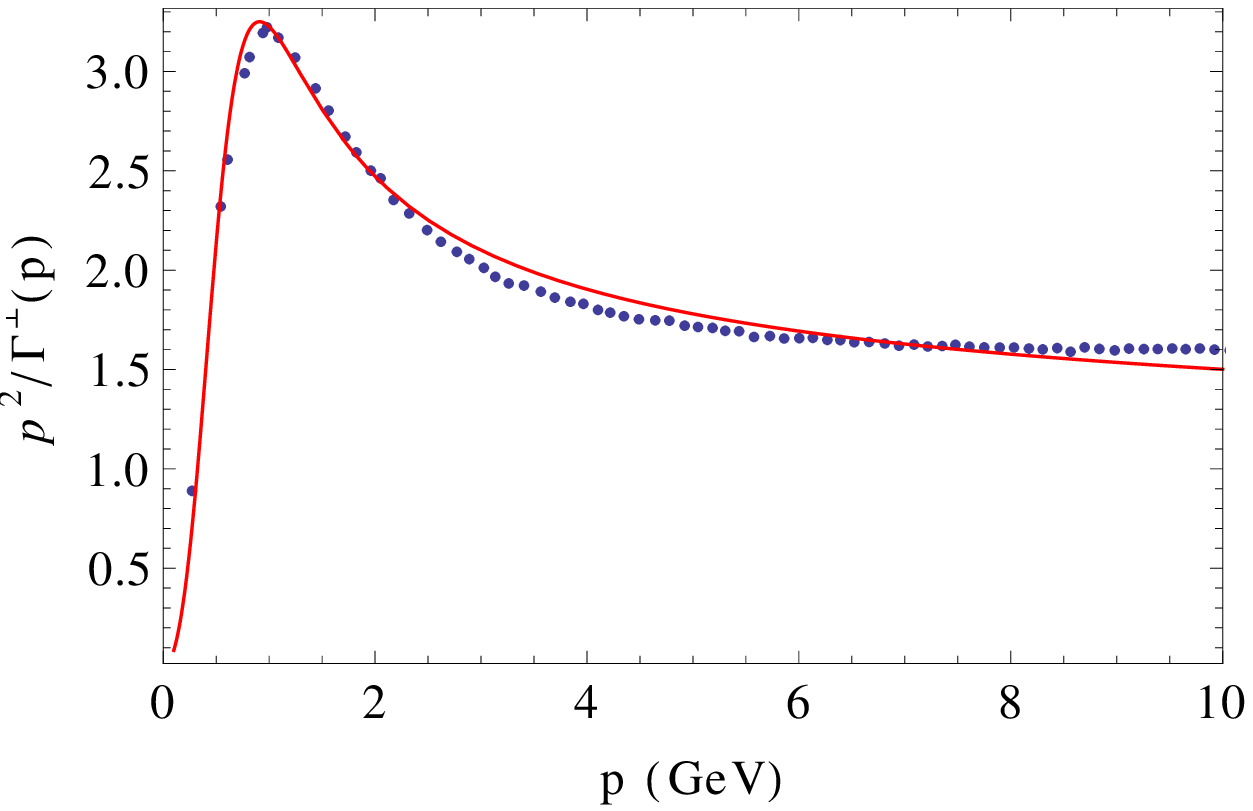}
 \includegraphics[width=\linewidth]{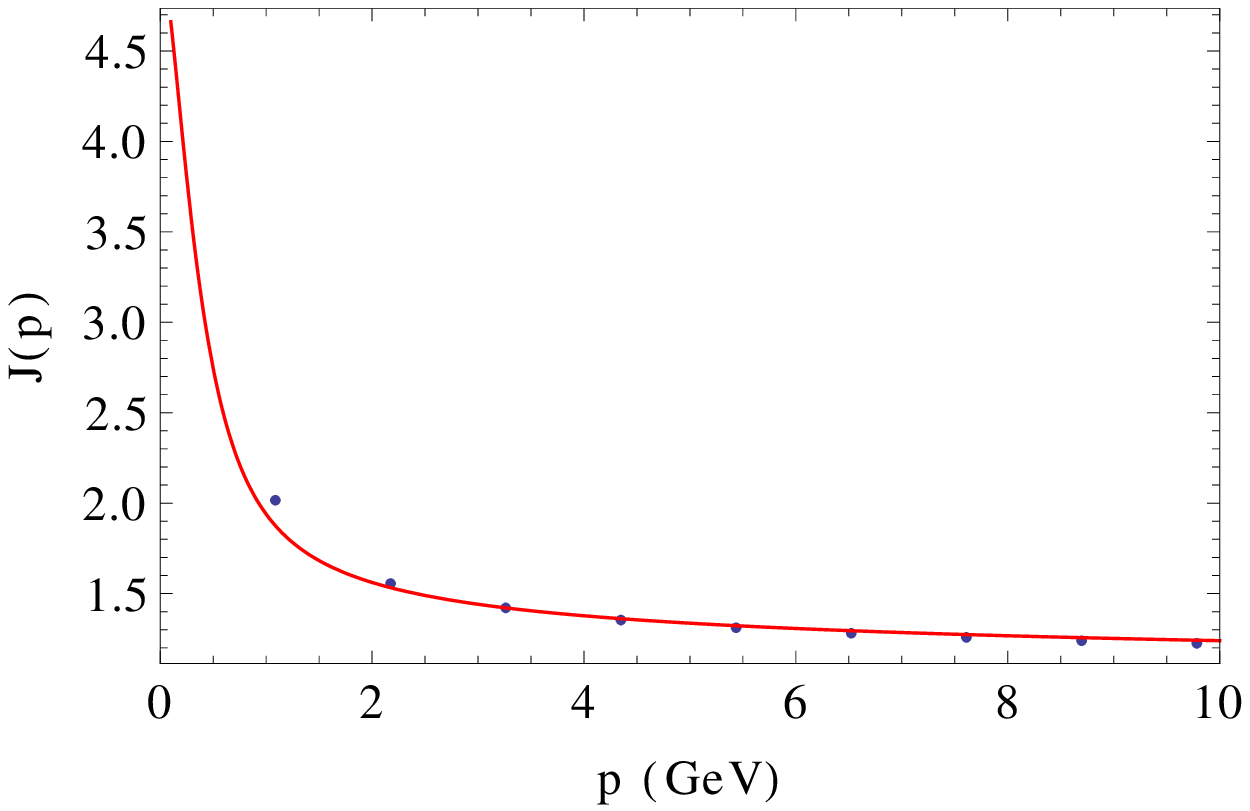}
\caption{Gluon dressing function (top) and ghost dressing function 
(bottom) as a function of momentum in $d=4$ for $N_f=2$. The points are lattice 
data of \cite{Sternbeck:2012qs}.}
\label{fig_propags_4dNf2}
 \end{figure}

\begin{figure}[htbp]
 \includegraphics[width=\linewidth]{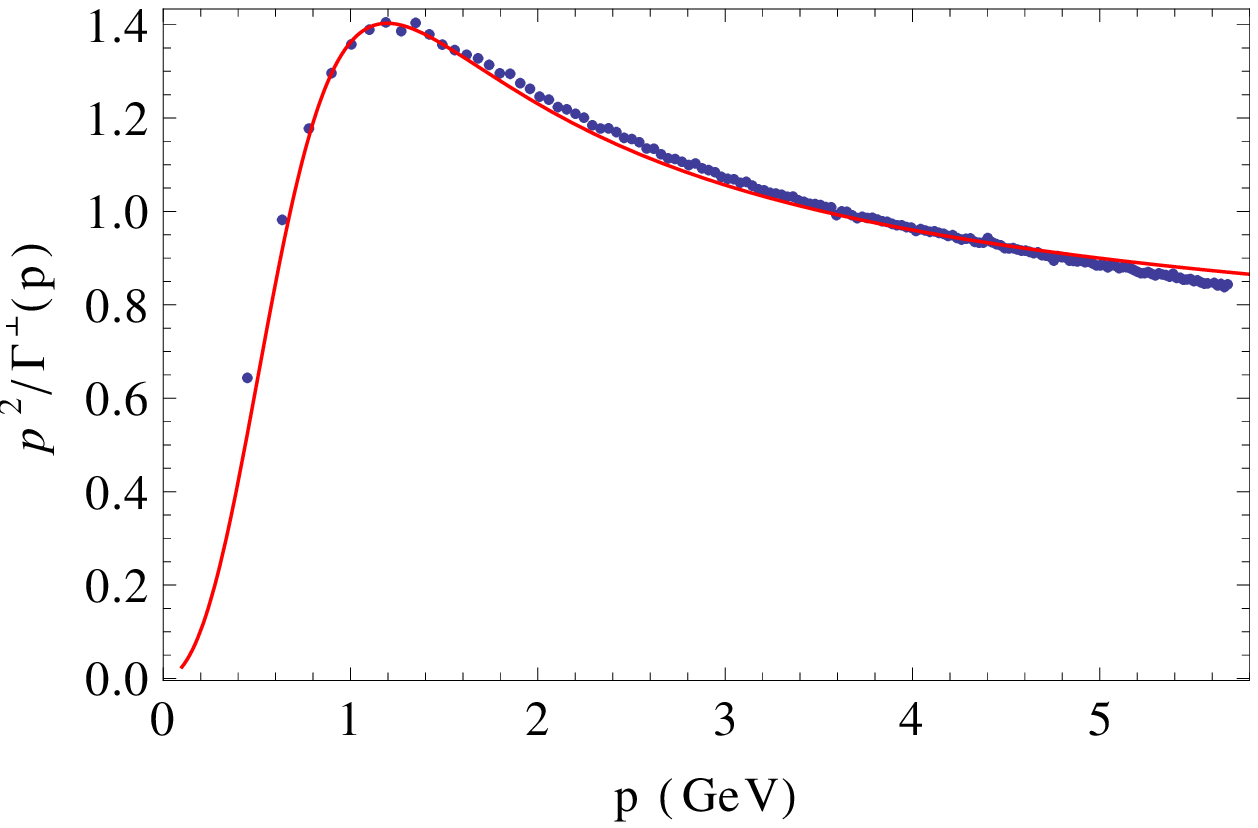}
 \includegraphics[width=\linewidth]{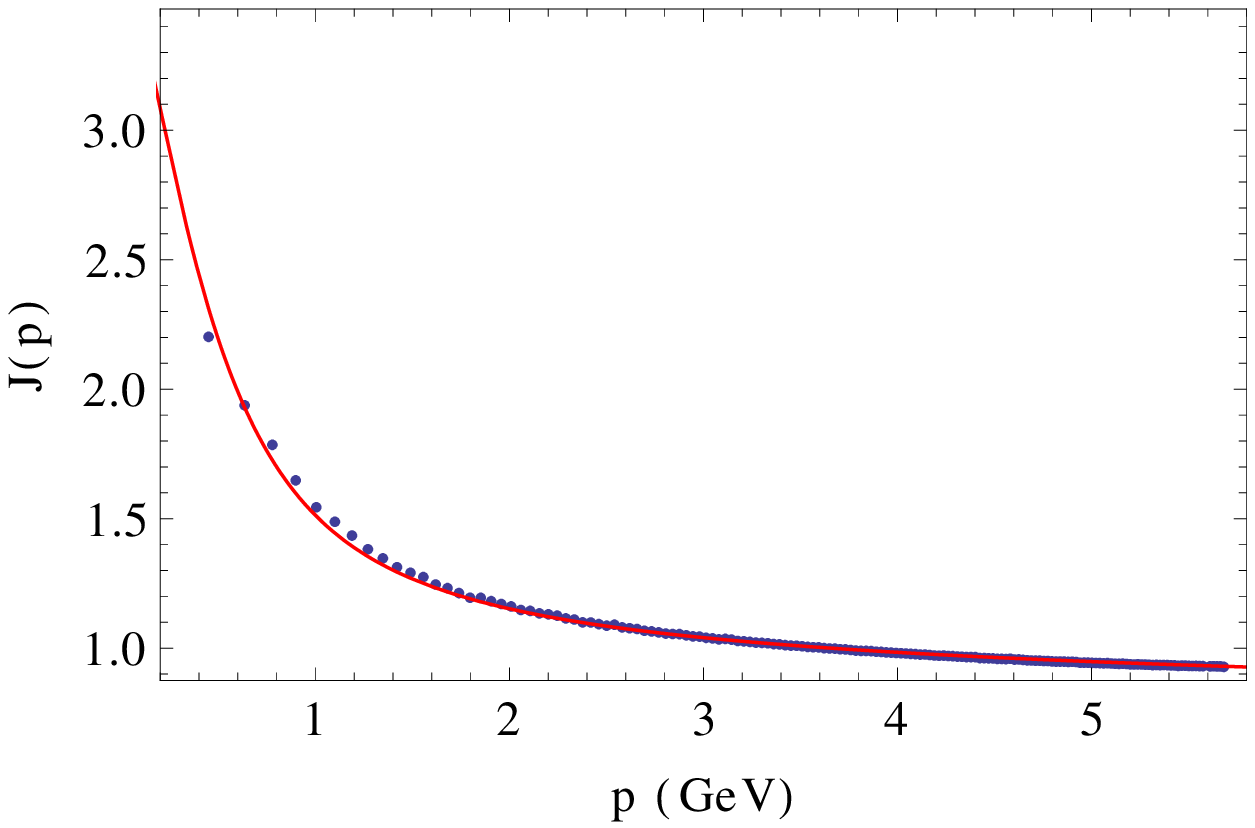}
\caption{Gluon dressing function (top) and ghost dressing function 
(bottom) as a function of momentum in $d=4$ for $N_f=2+1+1$. The points are lattice 
data of \cite{Ayala:2012pb}.}
\label{fig_propags_4dNf211}
 \end{figure}

For $N_f=2$, the best fits were obtained for $g =4.5$, $m=0.42 $~GeV
and $M_{u,d}=0.13 $~GeV. The corresponding gluon and ghost propagators are
depicted in Fig.~\ref{fig_propags_4dNf2}. For $N_f=2+1+1$, the best
fits were obtained for $g =5.3$, $m=0.56 $~GeV and $M_{u,d}=0.13
$~GeV. The corresponding gluon and ghost propagators are shown in
Fig.~\ref{fig_propags_4dNf211}. The comparison is very satisfactory,
with an error of at most three percent. Similar precisions were already
obtained in the quenched situation, see
\cite{Tissier:2010ts,Tissier:2011ey,Pelaez:2013cpa}.

It is important to stress that {\it a priori} we have three parameters to fix
but, in fact, in QCD, once the scale is fixed, the coupling constant should not
be arbitrary. At a given scale, its value should be compatible to the RG evolution
from the known value (for example, at the $Z^0$ mass scale). We verified that
the RG evolution from the coupling $g=5.3$ to the $Z^0$ mass scale gives a coupling
at that scale of $g=1.4$ to be compared to $g=1.2274$ (see, for example \cite{Blossier:2011tf}).
In order to arrive to this scale we had to use the bottom mass scale 4.2 GeV above where
we took into account a fifth quark in the running of the coupling.
This gives a 17\% error on the coupling that is more or less the overall 1-loop calculation.
This is similar to the corresponding calculation done in the quenched case \cite{Tissier:2011ey}.
In this sense, the coupling constant is not a free adjustable parameter. However, in order
to study the infrared, it is more convenient to fit the coupling at scales of the order of 1 GeV.

\begin{figure}[htbp]
 \includegraphics[width=4.2cm]{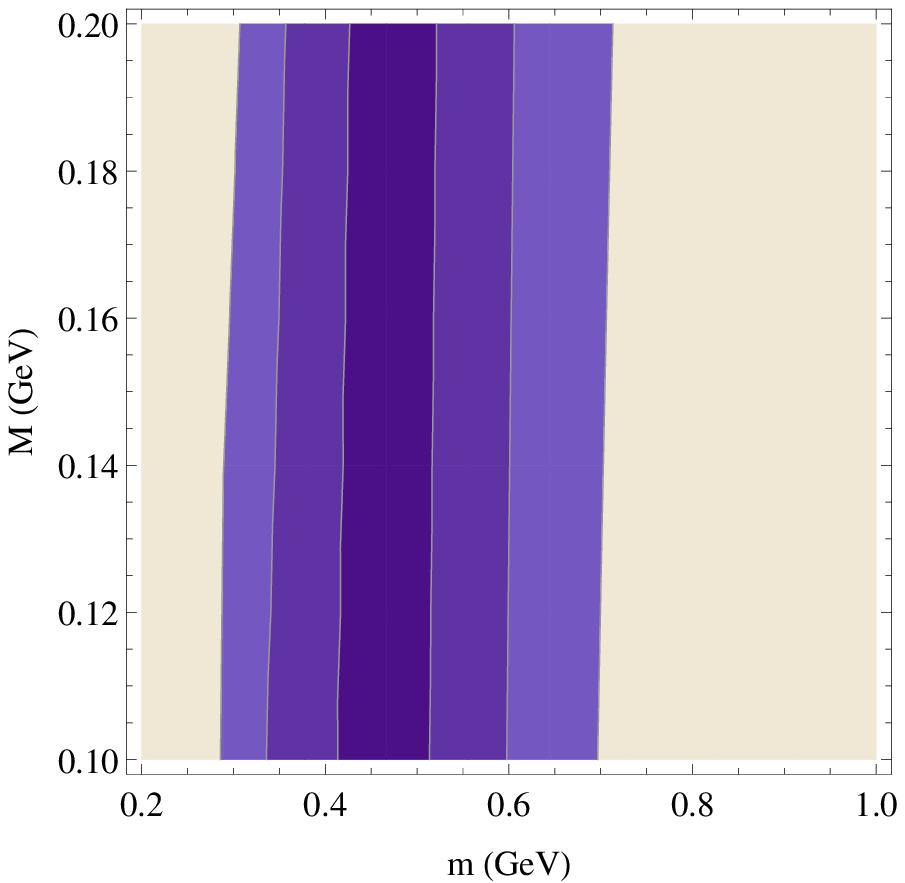}
 \includegraphics[width=4.2cm]{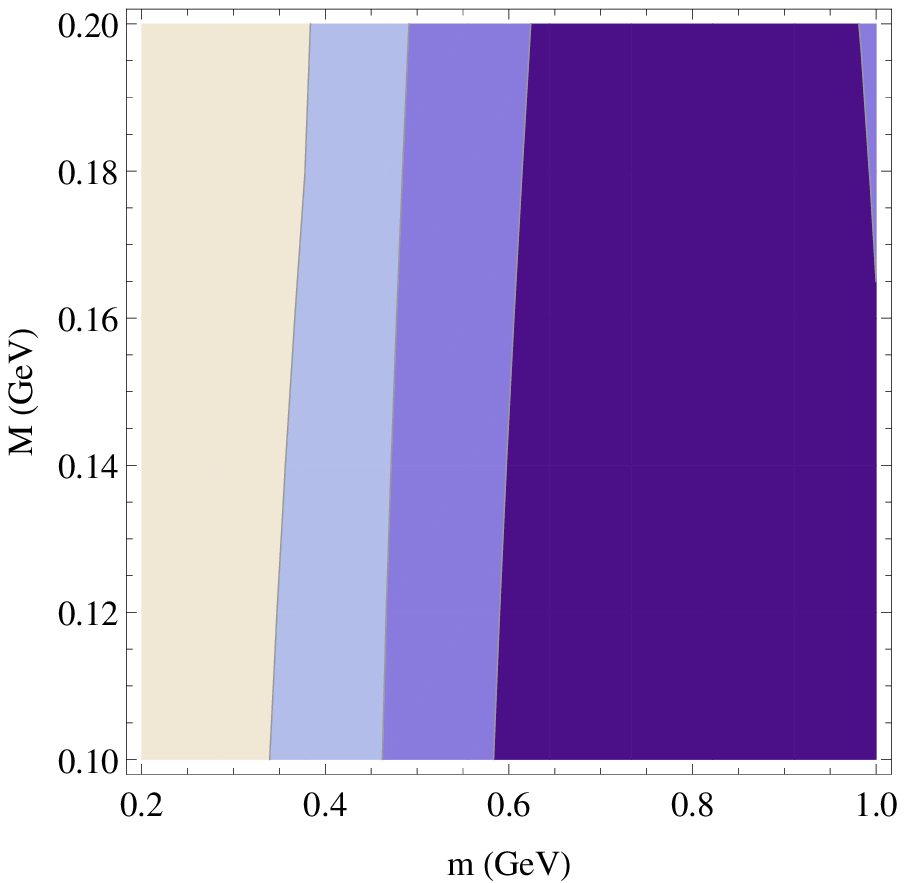}\\
 \includegraphics[width=4.2cm]{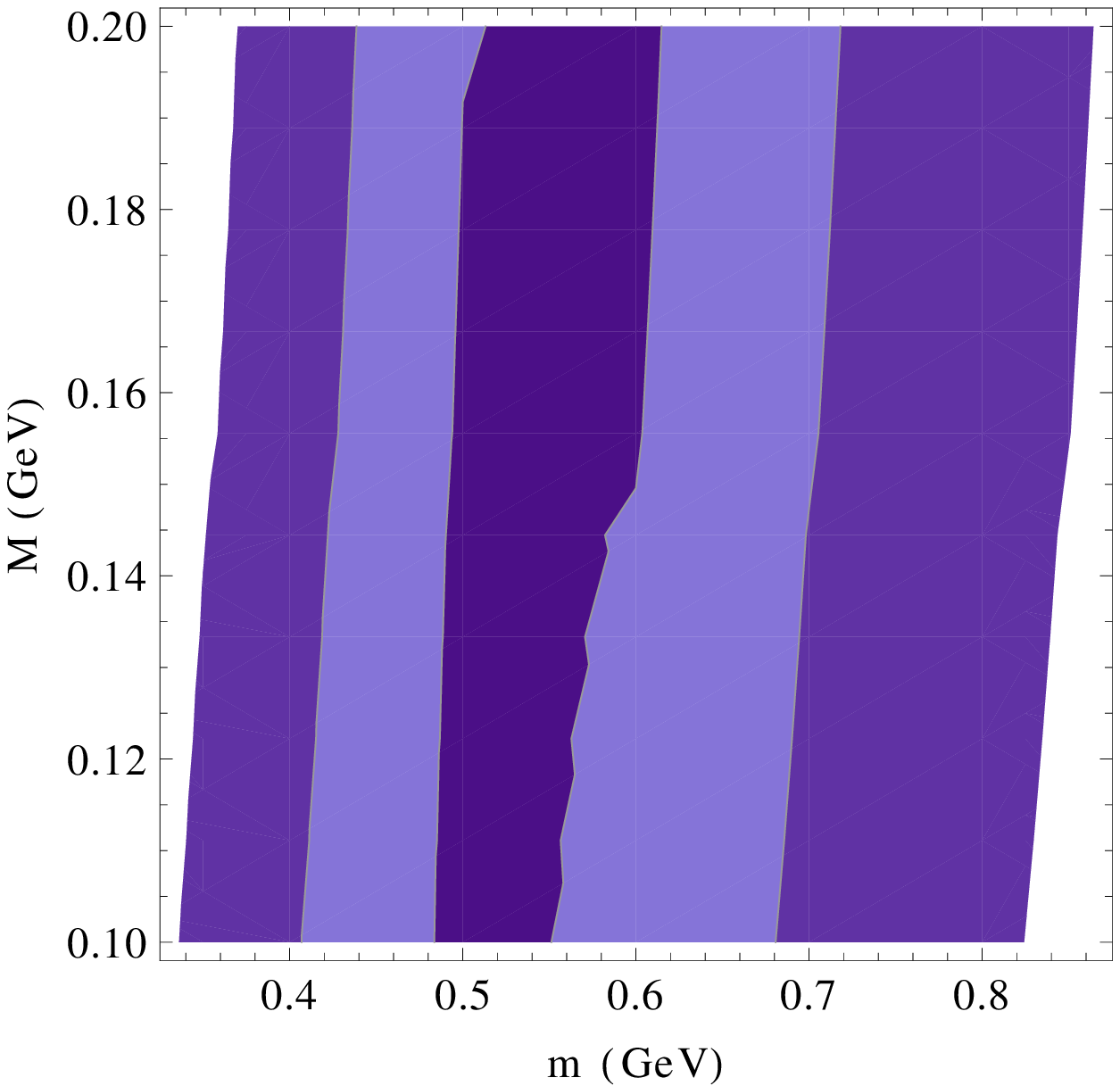}
 \includegraphics[width=4.2cm]{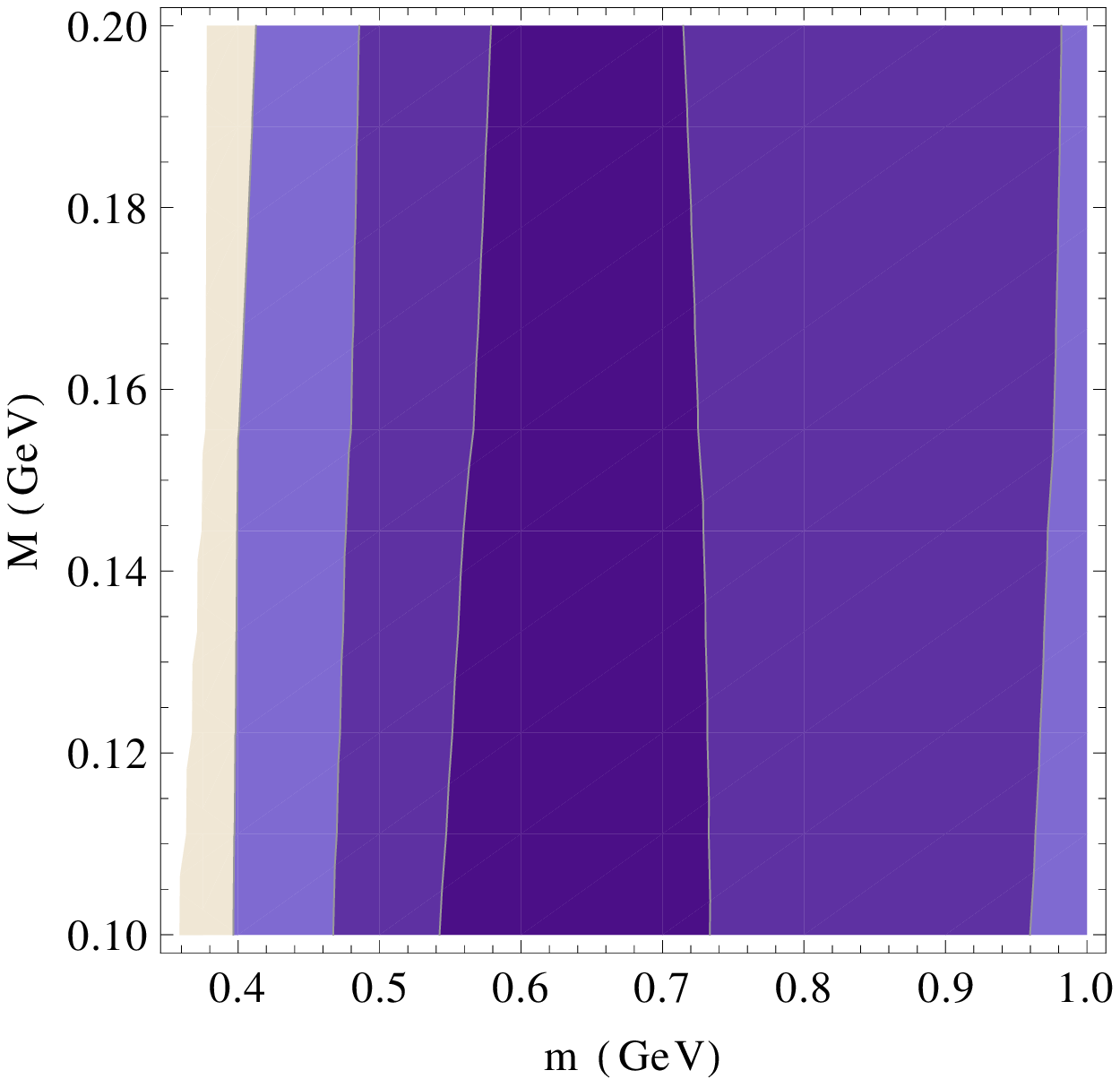}\\
 \caption{Contour levels for the quantities $\chi_{AA}$ (left) and
   $\chi_{c\bar c}$ (right) for $d=4$, both for $N_f=2$ (above) and
   $N_f=2+1+1$ (below). The contourlines correspond to 8\%, 10\% and
   12 \% (top left), 1.5\%, 2\% and 2.2 \% (top right), 3\%, 6\% and
   10 \% (bottom left), 1\%, 2\% and 4\% (bottom right).The quality
   of the fits is almost insensitive to the value of the quark mass.}
\label{errorcontourlevels}
 \end{figure}

 The contribution of the new diagram with an internal loop of quark is
 not strong enough to modify considerably the infrared behavior of the
 propagators due to the IR-safe structure of the quark propagator.
 Moreover the optimal values of the coupling constant and the gluon
 mass do not depend on slight changes in the quark mass.  That means
 that the error for the gluon and ghost propagators do not change
 significantly if we change the value of the light quark mass
 slightly. Fig.~\ref{errorcontourlevels} shows the contour regions for
 the errors.  The contour regions are
 vertical, showing that the quality of the fit is almost insensitive
 to the quark mass (vertical axes).
As expected, if the quark mass considerably increases at values of the
order of the GeV, the propagators tend to those obtained in the quenched approximation.

\begin{figure}[htbp]
 \includegraphics[width=\linewidth]{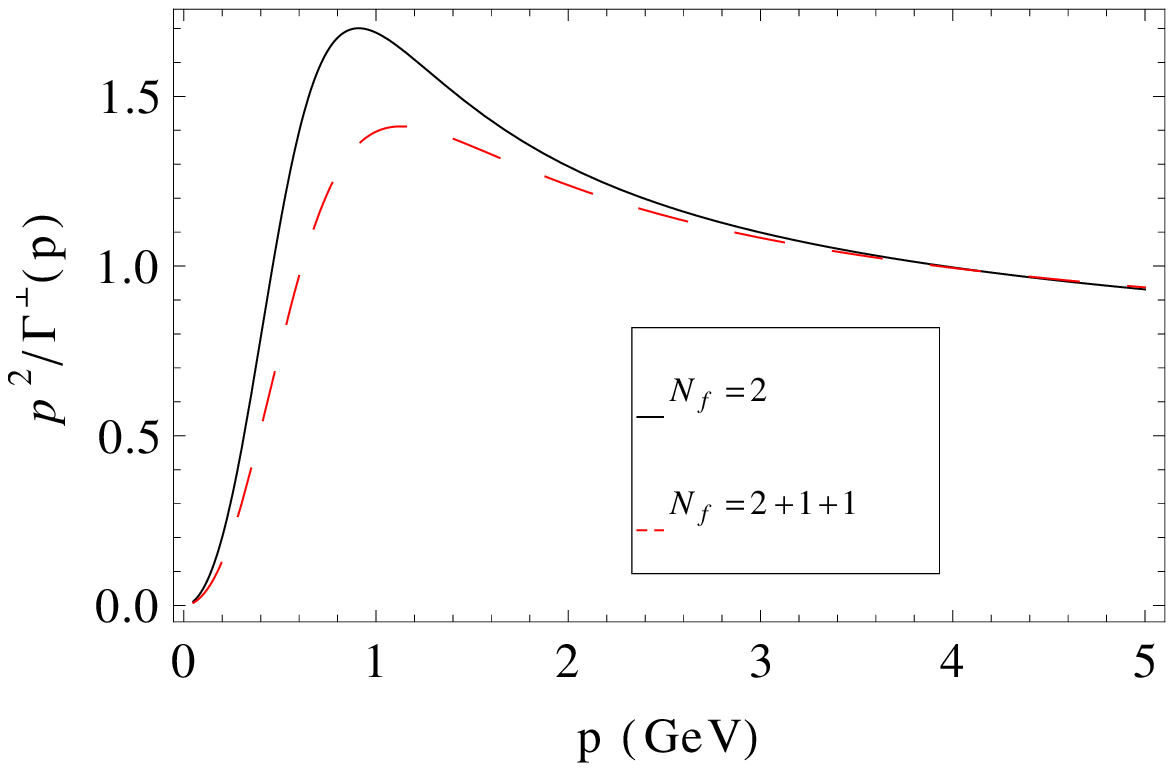}
 \includegraphics[width=\linewidth]{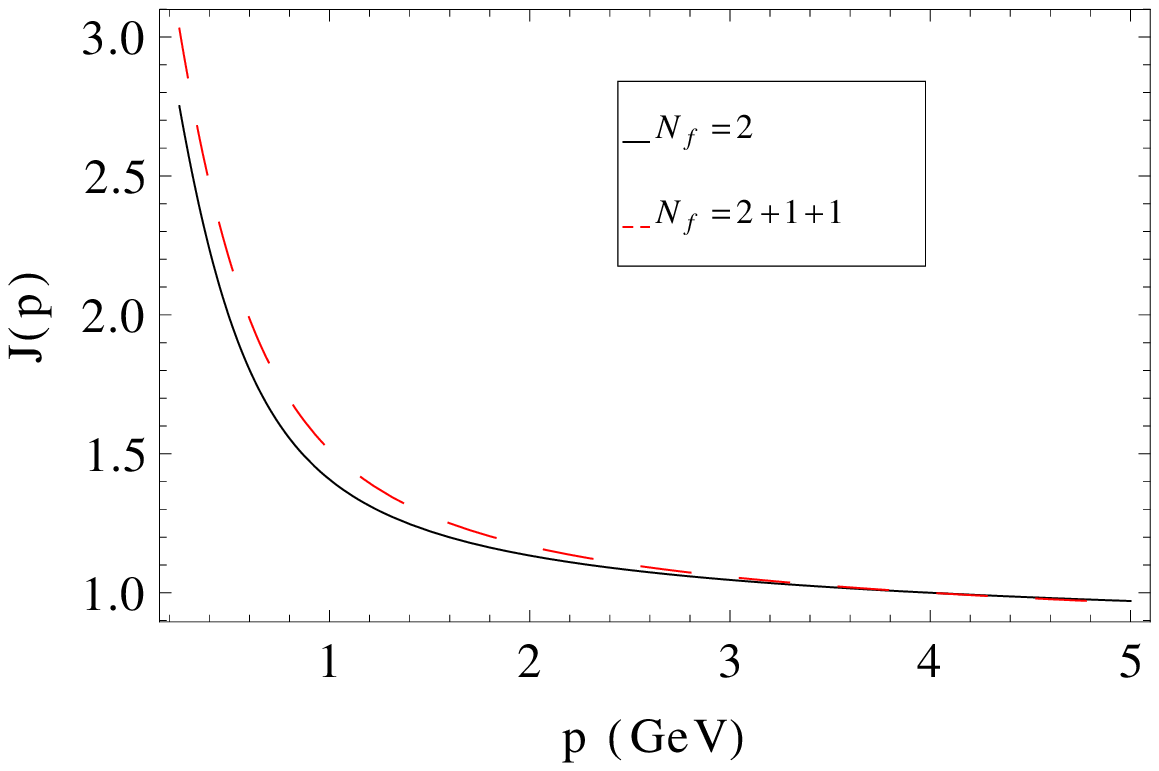}
 \caption{Gluon dressing function (top) and ghost
   dressing function (bottom) for different number of flavors in
   $d=4$}
\label{Nf}
 \end{figure}

It is interesting to compare the results obtained for the gluon
propagator and the ghost dressing function at one loop for different
number of quarks.  In Fig. \ref{Nf} we present the gluon propagator
and the ghost dressing function for $N_f=2$ and $N_f=2+1+1$. To
compare their infrared behaviors, we normalized the curves such that
they coincide at $4$ GeV. As was observed in lattice simulation
\cite{Ayala:2012pb,Bowman:2004jm}, the addition of heavy quarks leads
to a suppression of the IR saturation point \footnote{This effect is a
  consequence of both varying $N_f$ {\it and} changing the
  renormalization-group trajectory. If we change only $N_f$ but use
  the same initial conditions (at, say, $\mu_0=4$~GeV to obtain
  similar correlations in the UV), the quarks tend to suppress the
  gluon propagator around 1~GeV but enhance it at very low
  momenta.}. We also see that the ghost dressing function is enhanced.

\section{Results in the quark sector}
\label{sectionNf21}
We computed the quark propagator for arbitrary number of flavors but
as the only available data for the quark sector corresponds to two
degenerates light quarks plus a heavy one ($N_f=2+1$), we show the
results only for this case. As in the previous section, we fixed the ratio of the light
and heavy quarks to the ratio of the bare masses used in lattice simulations. Accordingly, in the present case, we fix the mass of
the heavy quark to be five times heavier than the light one at $\mu_0=1$~GeV.

As discussed in the previous section, the fits of the gluon and ghost
propagators are rather insensitive to the choice of the quark mass
(see Fig.~\ref{errorcontourlevels}). However our results for
$M_{u,d}(p)$ depend strongly on this parameter. Therefore we fixed the
fitting parameters by minimizing simultaneously the error on the gluon
propagator and on the function $M_{u,d}(p)$, $\chi_M$. (Note that the ghost
propagator was not extracted from lattice simulations for $N_f=2+1$
and could not be used for fixing the parameters, see below.) It is defined
as
\begin{equation}
 \chi^2_M=\frac{1}{N}\sum_i M^{-2}_{\rm th.}(\mu_0)
\left(M_{\rm lt.}(p_i)-M_{\rm th.}(p_i)\right)^2
\end{equation}
Note that we used a slightly different definition of the error. The reason is that the
mass function becomes rapidly close to zero when the momentum grows. Accordingly a
relative error point by point would be badly estimated because the signal to noise ratio
of lattice data becomes of order one. In order to avoid this problem we normalized the
data with a typical value of the mass in the infrared (at the scale $\mu_0$).

\begin{figure}[htbp]
 \includegraphics[width=\linewidth]{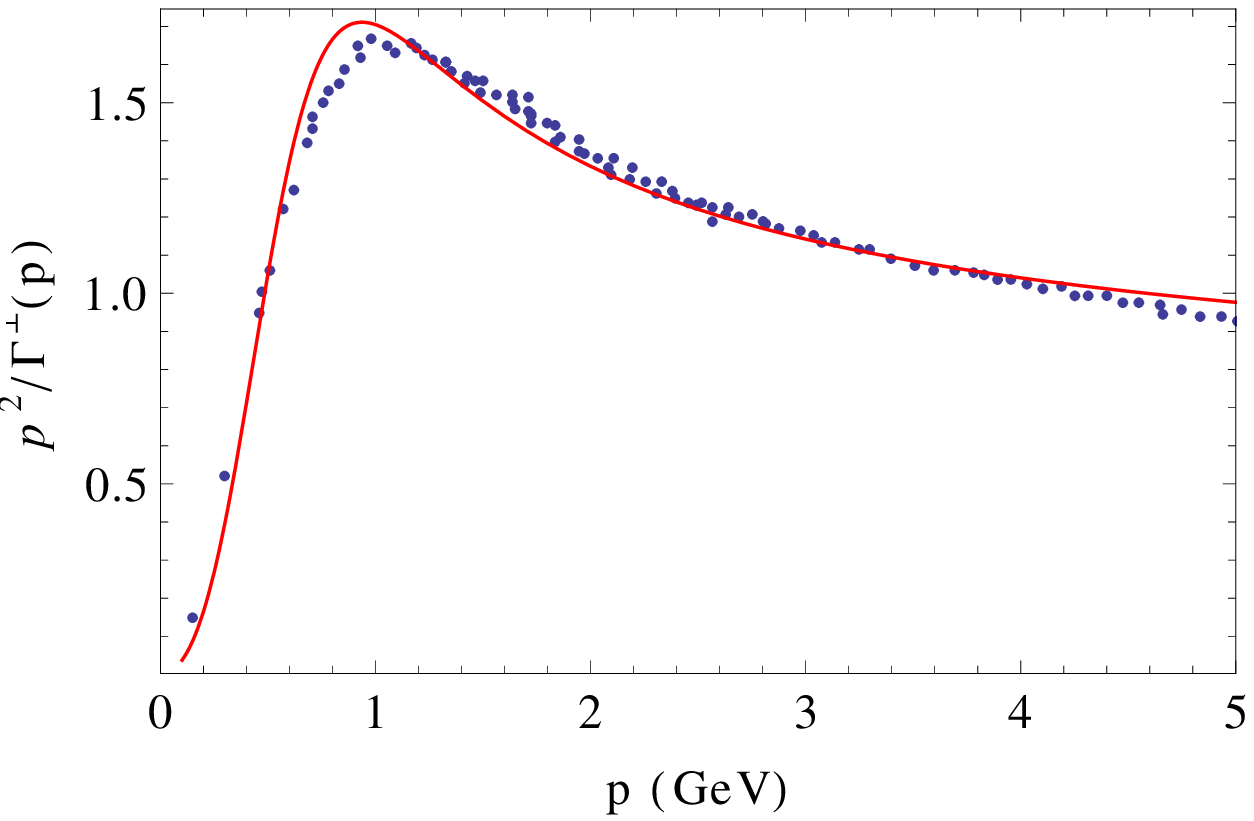}
 \includegraphics[width=\linewidth]{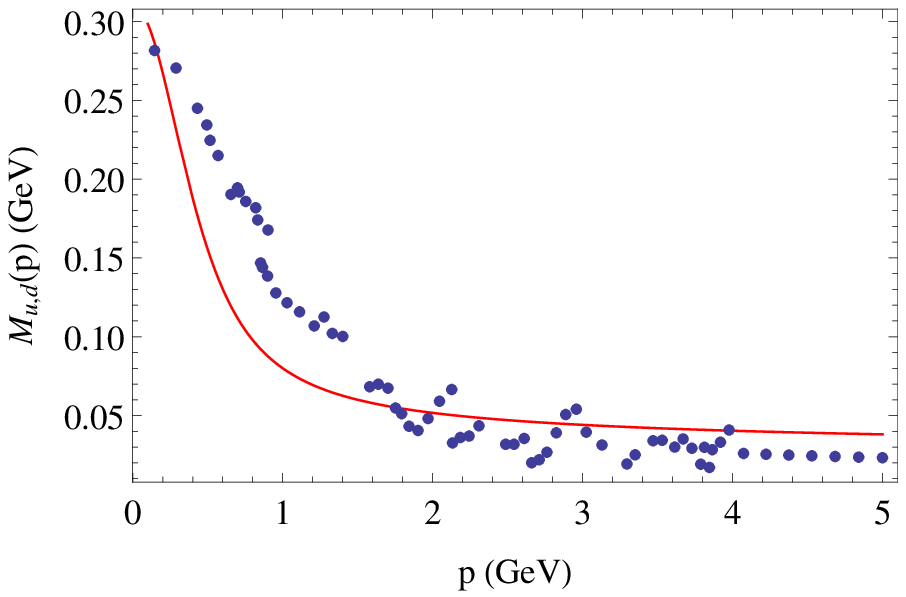}
\caption{Gluon dressing function (top) and quark mass $M_{u,d}(p)$ (bottom) in $d=4$ for $N_f=2+1$.
 The parameters are $g=4.8$, $m=0.42$ GeV and $M=0.08$. The points are lattice data of \cite{Bowman:2004jm,Bowman:2005vx}.}
\label{Nf21M08}
 \end{figure}

The best
fit parameters are $g=4.8$, $m=0.42 $~GeV and $M_{u,d}=0.08 $~GeV at
$\mu_0=1$~GeV. We insist on that the coupling and gluon mass are fixed from
the gluon propagator and once this is done, we fixed the mass parameter from the
scalar part of the quark propagator. The results obtained are shown in Fig.~\ref{Nf21M08}.
The gluon propagator is again reproduced with an accuracy similar to
that obtained for $N_f=2$ and $N_f=2+1+1$.

\begin{figure}[htbp]
 \includegraphics[width=4.8cm]{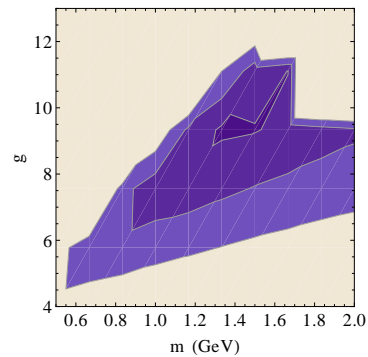}
\caption{Contour levels for the quantity $\chi_{M}$. The contourlines correspond to 10\%, 20\% and
   30 \%.}
\label{Nf21M14}
 \end{figure}

\begin{figure}[htbp]
 \includegraphics[width=\linewidth]{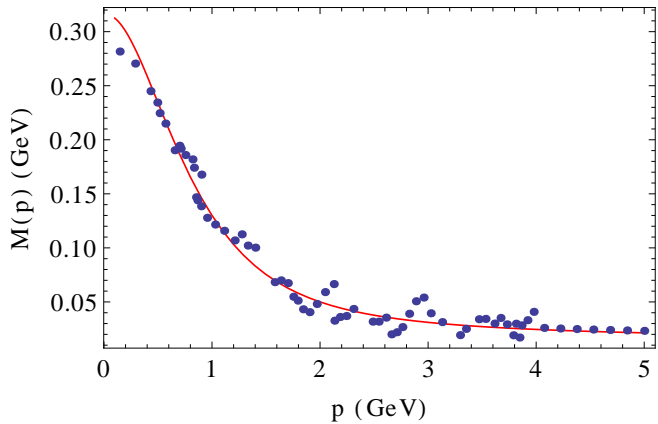}
\caption{Quark mass $M_{u,d}(p)$ in $d=4$ for $N_f=2+1$. The parameters are $g=9.6$, $m=1.4$ GeV and $M=0.13$ GeV.
The points are lattice data of \cite{Bowman:2004jm,Bowman:2005vx}.}
\label{Nf21M13}
 \end{figure}

\begin{figure}[htbp]
  \centering
  \includegraphics[width=\linewidth]{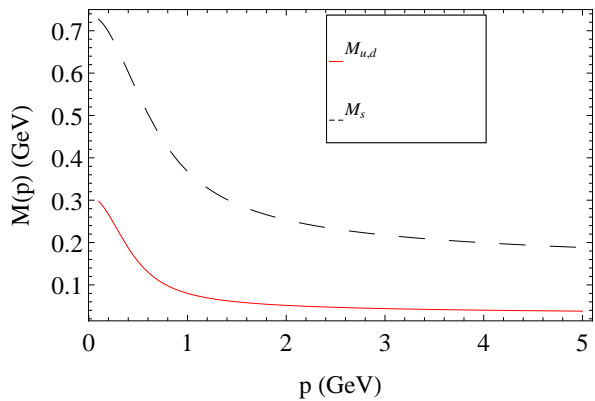}
  \caption{Quark masses of the light (full line, red) and heavy
    (dashed, black) quark.}
  \label{fig_mass_lh}
\end{figure}

The one-loop results for $M_{u,d}(p)$ compare correctly with lattice
simulations. In particular we retrieve the dramatic increase of the
mass at a momentum scale of the order of 2 GeV. We want to stress that we could
have a perfect fit of $M_{u,d}(p)$, but in doing so we would spoil the quality of
the gluon propagator. Instead, we choose to maintain the quality of the gluon
propagator at the prize of a not-so-good quark mass curve. In the Fig.~\ref{Nf21M14}
it is shown that there is a region of parameters where the fit is excellent (see  Fig.~\ref{Nf21M13}), but
we insist in using a value of the coupling compatible with a good fit for
the gluon dressing function. The ghost dressing
function was not extracted from lattice data and we can not test our
findings in this case. However, we expect that this function is rather
insensitive to the inclusion of a heavy quarks. Under this hypothesis,
it is interesting to compare our findings with the ghost dressing
function for $N_f=2+1+1$. We find fits (not shown) of the same quality
as in the previous section.
 For completeness, we also compare the mass functions for the light
 and heavy quarks in Fig.~\ref{fig_mass_lh}.

\begin{figure}[htbp]
 \includegraphics[width=\linewidth]{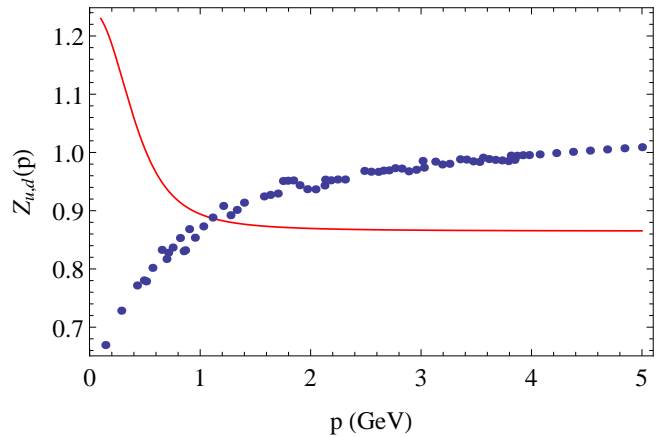}
\caption{$Z_{u,d}(p)$ in $d=4$. The points are lattice data of \cite{Bowman:2005vx}.}
\label{oneloopZNf21}
 \end{figure}

If the one-loop correlation functions compare very well for the
ghosts, gluons propagators and correctly for the function $M_{u,d}(p)$, the results for the
quark renormalization function $Z_{u,d}(p)$ is not of the same
quality~\footnote{For this reason, we did not use the function $Z(p)$
  for choosing the best fitting parameters, but concentrated on the
  function $M_{u,d}(p)$.}, as can be seen in
Fig. \ref{oneloopZNf21}. We attribute this mismatch to the fact that
the one-loop contribution to this function is unusually small. This is
a consequence of the fact that, when the mass of the gluon is zero,
$Z_{u,d}(p)$ has no contribution at one loop but it does have
contributions at two-loops. In this situation, the two-loop
contributions are not negligible and we do not expect a one-loop
calculation to be sufficient for describing the characteristics of the
field renormalization of the quarks.  In the next section, we estimate
the contribution of the two-loop diagrams and show that they have a
small impact on our findings for the ghost and gluon propagators and
on $M_{u,d}(p)$ but that they give a large contribution to
$Z_{u,d}(p)$.

\section{Preliminary estimate of two-loop contribution}
\label{sec_error}
The results presented in Fig.~\ref{oneloopZNf21} show that first order
perturbation theory is not enough to reproduce the behavior of the
$Z(p)$ in the Landau gauge.  For this reason we want to estimate the
importance of two loop corrections. Ideally, we should compute all the
two-loop diagrams and repeat the analysis performed above. Obviously
this work exceeds the purpose of the present article.

Instead we use a cartoon expression of the two-loop $\beta$ functions
(let us call it hybrid expression) obtained by complementing the
$\beta$ functions and anomalous dimensions derived here at one loop
with the ultraviolet two-loop contribution computed in
\cite{Gracey:2002yt} with a standard (massless) Faddeev-Popov
Lagrangian. To take into account the suppression of massive
contribution at low momenta, the ultraviolet two-loop contributions
have to be appropriately regularized in that regime~\footnote{It is
  worth to notice that even though the ghost propagator is massless
  all the two-loop diagrams are frozen in the infrared.}.  Here we
choose to modify the anomalous dimension presented in
\cite{Gracey:2002yt} in a minimal way by multiplying them by:
\[\frac{\mu^2}{\mu^2+s^2}\]
This factor goes to one in the UV limit where the whole expression matches 
with \cite{Gracey:2002yt}. On the other hand, in the infrared regime
it goes to zero as $\mu^2$, as expected. In the following, we present
our results for $s=1$~GeV but we checked that we obtain the same
conclusions by changing the parameter in the range 0.5---2 GeV.

\begin{figure}[htbp]
 \includegraphics[width=\linewidth]{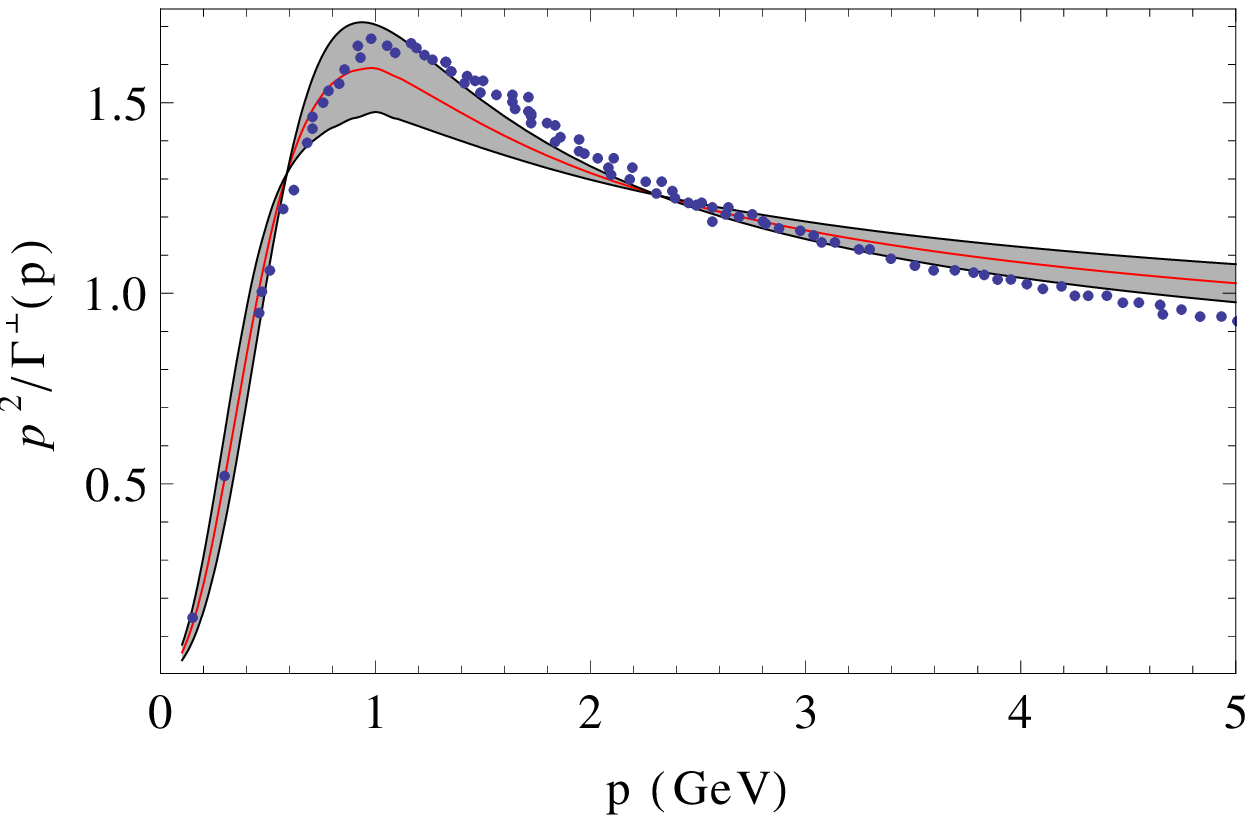}
 \includegraphics[width=\linewidth]{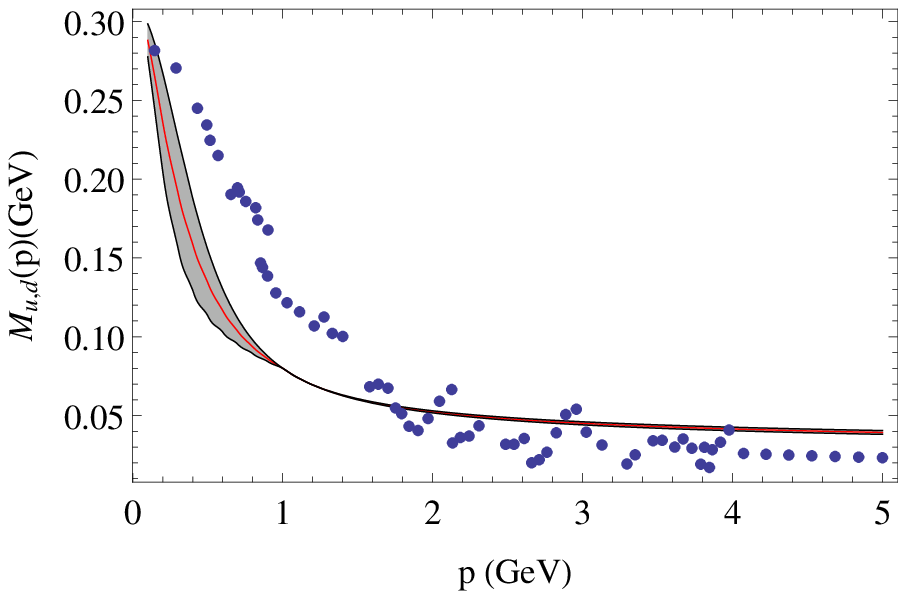}
 \includegraphics[width=\linewidth]{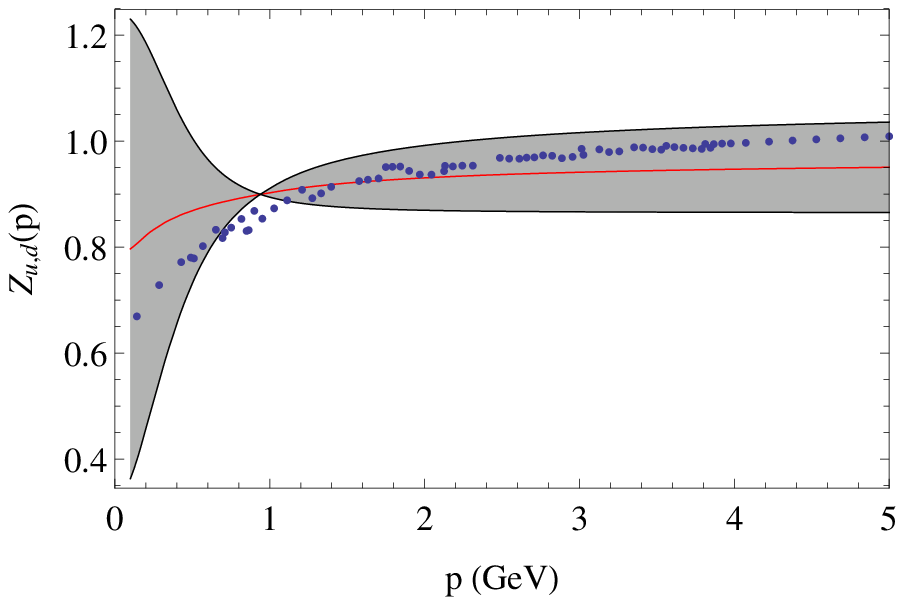}
 \caption{Gluon dressing function (top), quark mass
   $M_{u,d}(p)$ (middle) and $Z_{u,d}(p)$ (bottom) as a function of
   momentum in $d=4$. In red the hybrid calculation results (see text) and in black the
   one loop calculation and the symmetrized one loop results with respect to hybrid calculation. The points are lattice data of
   \cite{Sternbeck:2012qs} and \cite{Bowman:2004jm,Bowman:2005vx}.}
\label{zonas}
 \end{figure}

We have performed a fit of the lattice data with this hybrid flow
equations. The best fits were obtained for $g=3.4$, $m=0.4 $~GeV and
$M=0.08 $~GeV, again defined at $\mu_0=1$~GeV. These fits are depicted
in Fig.~\ref{zonas}, in the center of the shaded areas. We estimate
the error coming from neglecting the 2-loop contribution as the difference between the best fit in the hybrid model
and in the purely one-loop calculation. In the Fig.~\ref{zonas} the hybrid calculation
is taken as the central value (that includes a rough estimate of the two-loop contributions) and
the error band estimated as explained just before is represented by the shaded area. We clearly see
that higher corrections for the gluon propagator are small and
compatible with the discrepancy between our one-loop results and the
lattice data. The two-loop contribution to the function $M(p)$ is
small but does not seem sufficient to make the one-loop results
compatible with lattice data. This could be related with the
spontaneous breaking of chiral symmetry which was not treated here.
Finally, the corrections to $Z(p)$ are large and can explain the
discrepancy between the lattice data and the 1-loop results. As
mentioned in the previous section, this last function is therefore
more difficult to describe and it is necessary to go to next to
leading order to find a satisfactory matching.

\begin{figure}[htbp]
 \includegraphics[width=\linewidth]{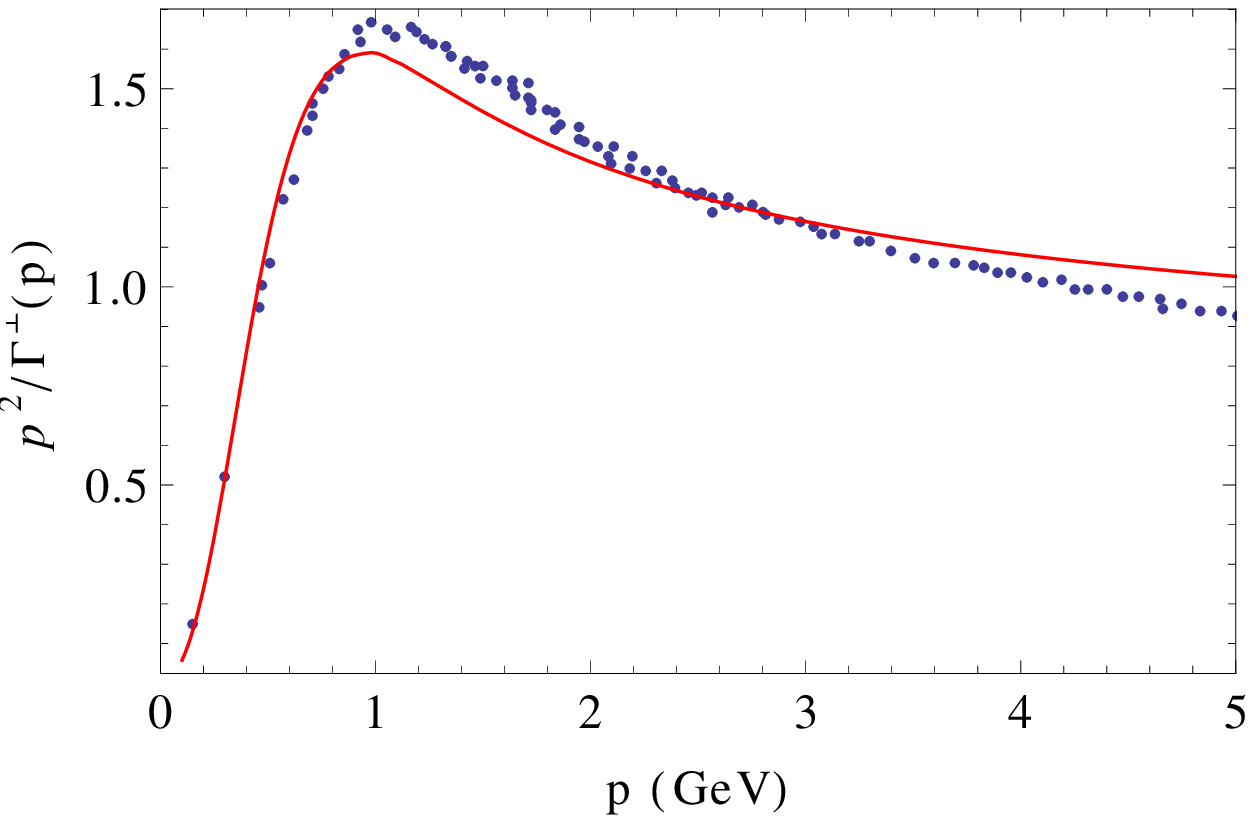}
 \includegraphics[width=\linewidth]{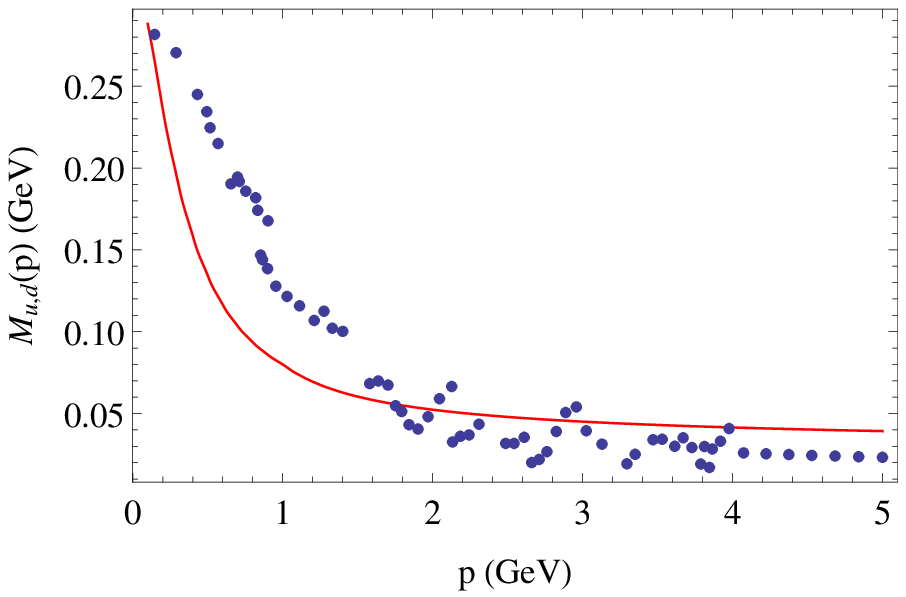}
 \includegraphics[width=\linewidth]{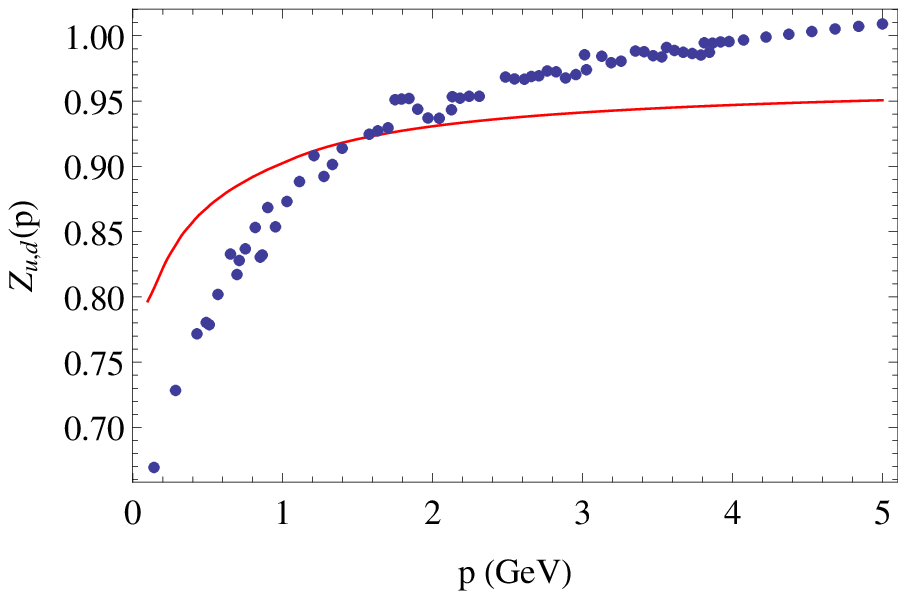}
\caption{Gluon dressing function (top), quark mass $M(p)$(middle) and $Z(p)$ (bottom) in $d=4$ using the hybrid approximation. The points are lattice data of \cite{Bowman:2004jm,Bowman:2005vx}.}
\label{fig_propagsGracey_4d}
 \end{figure}

\section{Conclusions}
In this article, we presented a one-loop calculation of the two-point
correlation functions of QCD in the Landau gauge. The effect of the
Gribov copies is minimally encoded in a mass term for the gluons. We
find that the gluon and ghost propagators are reproduced with high
precision. We estimate the higher order corrections to be small which
seems to indicate that the Curci-Ferrari model reproduces well the
lattice data.

On the other hand, one loop calculations are not enough to describe
properly all the properties of the quark sector. The one loop
contribution to $Z_{u,d}(p)$ function is very small and it vanishes if
the gluon mass goes to zero in Landau gauge.  That is why the two
loops contributions are important and we have to take them into
account if we want to quantitatively reproduce the $Z_{u,d}(p)$
function behavior.

The mass of the quark is correctly reproduced even though the accuracy
is not as good as it is for the gluon sector. It is even unclear if a
perfect matching for this quantity by such a simple one-loop
calculation is possible given that we have not studied the influence
of the chiral symmetry breaking. This analysis remains for a future
work.

As we discuss in the article, computing the two loop contribution can
help to estimate the validity of perturbation theory in this
model. That is why we are considering in doing the two loop
calculation completely. 

\begin{acknowledgments}
  The authors want to thank A.~Sternbeck, J.~Rodriguez-Quintero and
  B.~El-Bennich for kindly making available the lattice data.  We also
  want to thank J.~A.~Gracey for providing the two loop massless
  results for each diagram independently.  The authors want to
  acknowledge partial support from PEDECIBA and ECOS
  programs. M. Peláez wants to thank the ANII for financial support.

\end{acknowledgments}

\end{document}